\documentclass[lettersize,journal]{IEEEtran}
\usepackage{bm}
\usepackage{subfigure}
\usepackage{cite}
\usepackage{amsmath,amssymb,amsfonts}
\usepackage{graphicx}
\usepackage{textcomp}
\usepackage{xcolor}
\usepackage{algorithm}
\usepackage{algpseudocode}
\usepackage{float}
\usepackage{hyperref}
\usepackage{titlesec}
\usepackage{enumitem}

\def\BibTeX{{\rm B\kern-.05em{\sc i\kern-.025em b}\kern-.08em
    T\kern-.1667em\lower.7ex\hbox{E}\kern-.125emX}}
\begin{document}

\title{Channel-Aware Conditional Diffusion Model for Secure MU-MISO Communications}

\author{Tong Hui,
        Xiao Tang,
        Yichen Wang,
        Qinghe Du,
        Dusit Niyato,
        and Zhu Han
\thanks{T. Hui is with Shenzhen Research Institute of Northwestern Polytechnical University, Shenzhen 518057, China, and also with the School of Electronics and Information, Northwestern Polytechnical University, Xi'an 710072, China. (e-mail: htnwpu@mail.nwpu.edu.cn)}
\thanks{X. Tang is with the School of Information and Communication Engineering, Xi'an Jiaotong University, Xi'an 710049, China, and also with Shenzhen Research Institute of Northwestern Polytechnical University, Shenzhen 518057, China. (e-mail: tangxiao@xjtu.edu.cn)}
\thanks{Y. Wang and Q. Du are with the School of Information and Communication Engineering, Xi'an Jiaotong University, Xi'an 710049, China. (e-mail: wangyichen0819@mail.xjtu.edu.cn, duqinghe@mail.xjtu.edu.cn)}
\thanks{D. Niyato is with the College of Computing and Data Science, Nanyang Technological University, Singapore. (e-mail: dniyato@ntu.edu.sg)}
\thanks{Z. Han is with the Department of Electrical and Computer Engineering, University of Houston, Houston 77004, USA, and also with the Department of Computer Science and Engineering, Kyung Hee University, Seoul 446-701, South Korea. (e-mail: hanzhu22@gmail.com)}
}
\maketitle

\begin{abstract}

While information security \!is a fundamental requirement for wireless communications, conventional optimization-based 
approaches often struggle with real-time implementation, and deep models, typically discriminative in nature, may lack the 
ability to cope with unforeseen scenarios.  To address this challenge, this paper investigates the design of legitimate beamforming and artificial noise (AN) to achieve physical layer security by exploiting the conditional diffusion model. Specifically, we reformulate  the security optimization as a conditional generative process, using a diffusion model to learn the inherent distribution of near-optimal joint beamforming and AN strategies. We employ a U-Net architecture with cross-attention to integrate channel state information, as the basis for the generative process.  Moreover, we fine-tune the trained model using an objective incorporating the sum secrecy rate such that the security performance is further enhanced. Finally, simulation results validate the learning process convergence and demonstrate that the proposed generative method achieves superior secrecy performance across various scenarios as compared with the baselines.
\end{abstract}


\section{Introduction}
With the rapid growth of the data volume in emerging sixth-generation (6G) wireless networks, protecting confidential information from eavesdropping has become a key challenge. Due to the low-complexity implementation and potential of perfect secrecy, the physical layer security (PLS) technology has raised wide attentions. In particular, multi-antenna beamforming, often combined with artificial noise (AN) injection, utilizes spatial degrees of freedom to enhance legitimate transmission while simultaneously reducing potential wiretap channels, thereby providing an effective means of protecting the secrecy~\cite{8368140}. Therefore, PLS techniques provide an effective and scalable approach for secure communications in the dense, dynamic, and heterogeneous wireless environments envisioned for 6G networks.

However, when extending the PLS techniques to multi-user scenarios, the transmission strategy design needs to incorporate inter-user interference, variable channel conditions and scenario characteristics into consideration. Traditional optimization methods primarily achieve joint design through approximation and iteration~\cite{7605477,9973364}, often yielding suboptimal results unsuitable for online inference. Consequently, recent work resorts to data-driven methods for PLS strategy design. For instance, multilayer perceptrons (MLPs) facilitate real-time interference management and security enhancement in wiretap channels~\cite{8444648,9663220}. Similarly, graph neural networks achieve near-optimal multi-user multiple-input multiple-output (MU-MIMO) precoding~\cite{10345473}, and secure beamforming through message-passing mechanisms~\cite{10955456}. These deep learning-based approaches, being fundamentally discriminative models, classify interference and legitimate signals by learning the decision boundaries directly from observed channel features. As a result, they are usually sensitive to noise and perturbations, and their robustness degrades when channel statistics or network topology vary to unseen scenarios.

Meanwhile, diffusion model has opened up new avenues for the design of wireless communication strategies by modeling complex high-dimensional data distributions~\cite{NEURIPS2020_4c5bcfec}. 
Compared to generative adversarial networks and variational autoencoders, which may be constrained by unstable training, mode collapse, or reconstruction blurring, this generative artificial intelligence method offers potential advantages in achieving high-fidelity sample generation, thereby attracting wide interest. 
Conditional diffusion models~\cite{9878449} have demonstrated robust adaptability to channel uncertainty and interference in channel denoising, deviceless human detection and arrival angle estimation~\cite{10480348,10557650,10599123}. Unlike discrimination models, diffusion model inherently learns the latent conditional distribution of the optimal solution. For security-oriented applications, by iteratively removing injected noise, the diffusion model restores the secure transmission strategies from noisy observations, thus making it particularly effective in the uncertain and interference-limited scenarios.  Moreover, compared with traditional deep learning methods that optimize for single-point solutions, this generative methord produces multiple high-performance candidate strategies, enhancing the robustness of wireless communications against eavesdropping attacks.

Consequently, this paper proposes a channel-aware conditional diffusion model for multi-user multiple-input single-output (MU-MISO) secure communications. We consider the joint beamforming and AN optimization within a conditional generation perspective. To this end, we first construct a dataset of high-quality beamforming strategy under different channel realizations through numerical optimization methods. During training, the diffusion model learns to denoise Gaussian noise input into secure strategy vectors by capturing the conditional distribution inherently within the dataset. Subsequently, fine-tuning is performed by explicitly incorporating the secrecy rate as the ultimate optimization objective, enabling rapid convergence while further exploring high-performance regions of the strategy solution space. Simulation results show that the proposed generative method has superior secrecy performance in various scenarios compared with the baselines.
\section{System Model and Problem Formulation}

We consider a downlink MU-MISO system in which a base station (BS) equipped with \(M\) antennas communicates with \(K\) single antenna legitimate users. There are \(L\) single-antenna eavesdroppers wiretapping on legitimate transmissions. Let \(s_k\) be the symbol sent to user \(k\). The BS employs beamforming vector \(\bm{w}_k\!\in\!\mathbb{C}^{M}\)\! to transmit \!\(s_k\), while simultaneously injecting AN. The composite transmit vector is
\begin{equation}
\bm{x} = \sum_{k \in \mathcal{K}} \bm{w}_k^{\dagger} s_k + {\bm{V}} \!\bm{\eta},
\end{equation}
where \!\({\bm{V}}\!\in\!\mathbb{C}^{M \times J}\) is the AN precoding matrix composed of $J$ column vectors and \({\bm{\eta}}\!\sim\!\mathcal{CN}\left(0,\bm{I}_J\right).\) The resulting AN vector thus follows \!\({\bm{V}}\!\bm{\eta}\!\sim\!\mathcal{CN}\!\left(0,\!P \bm{V} \!\bm{V}^{\dagger}\right).\) Under the constraint of the total power $P$, the beamforming and AN vectors must satisfy
\begin{equation}
    \setlength{\abovedisplayskip}{3pt}
\setlength{\belowdisplayskip}{2pt}
P \sum_{{k \in \mathcal{K}}} \| \bm{w}_k \|^2 + P \sum_{{j \in \mathcal{J}}} \| \bm{v}_j \|^2 \leq P,
\end{equation}
where \(\bm{v}_j\) is the \(j\)-th column of \({\bm{V}}\).

Let \(\bm{h}_k\!\in\!\mathbb{C}^{M}\) represent the channel vector from the BS to the \(k\)-th user, and \(\bm{h}_{\mathrm{E},l}\!\in\!\mathbb{C}^{M}\) to represent 
the channel vector from the BS to the \(l\)-th eavesdropper. We assume that all channels are independent of 
each other and that the channel gain remains constant over the transmission block duration. The BS has 
perfect channel state information (CSI).Without loss of generality, we treat multi-user interference as noise and assume unit bandwidth, yielding the achievable communication rate \(R_k\) for legitimate user-\(k\) is

\begin{equation}
    \setlength{\abovedisplayskip}{2pt}
    \setlength{\belowdisplayskip}{2pt}
R_k\!=\!\log_2\!\left(\!
    1\!+\!\frac{
    \left| \bm{h}_k^{\dagger} \bm{w}_k \right|^2
    }{
    \sum_{\substack{i \in \mathcal{K} \setminus \{k\}}}\!\!
    \left| \bm{h}_k^{\dagger} \bm{w}_i \right|^2\!\!+\!\sum_{\substack{j \in \mathcal{J}}}
    \left| \bm{h}_k^{\dagger} \bm{v}_j \right|^2\!\!+\!\sigma_0^2 / P
    }
    \right),
\end{equation}
where $\sigma_0^2$ is the thermal-noise power. 
Similarly, the wiretap rate of the message \(s_k\) obtained from the eavesdropper-\(l\) is
\begin{equation}
    \setlength{\abovedisplayskip}{2pt}
    \setlength{\belowdisplayskip}{2pt}
R_{l,k}^{\mathrm{e}}\!=\!\log_2\!\left(\!
    1\!+\!\frac{
    \left| \bm{h}_{\mathrm{E},l}^{\dagger} \bm{w}_k \right|^2
    }{
    \displaystyle\sum_{\substack{i \in \mathcal{K} \setminus \{k\}}}\!\!
    \left| \bm{h}_{\mathrm{E},l}^{\dagger} \bm{w}_i \right|^2\!\!+\!\displaystyle\sum_{\substack{j \in \mathcal{J}}}
    \left| \bm{h}_{\mathrm{E},l}^{\dagger} \bm{v}_j \right|^2\!\!+\!\sigma_\mathrm{0}^2 / P
    }
    \right).
\end{equation}

Considering that multiple eavesdroppers can decode the transmission information of all users, the secrecy rate for 
user-\(k\) is determined by the worst-case wiretap channel. Therefore, the secrecy rate that can be achieved by user-\(k\) is
\begin{equation}
    \setlength{\abovedisplayskip}{2pt}
\setlength{\belowdisplayskip}{2pt}
R_k^{\mathrm{sec}} = \left( R_k - \max_{l \in \mathcal{L}}\{R_{l,k}^{\mathrm{e}}\} \right)^{+},
\end{equation}
where \!\((\cdot)^{+}\! = \!\max(\cdot, 0)\). However, directly handling the non-differentiable max operation is challenging for gradient-based optimization methods, such as training neural networks. Thus, we introduce a differentiable approximation for the max function using the LogSumExp operation as 
\begin{equation}
    \setlength{\abovedisplayskip}{2pt}
\setlength{\belowdisplayskip}{2pt}
    R_k^{\mathrm{sec}} \approx  \left(  R_k -\alpha \log \left(\sum_{{l \in \mathcal{L}}} \exp\left( \frac{R_{l,k}^{\mathrm{e}}}{\alpha} \right) \right) \right)^{+},
\label{eq6}    
\end{equation}
where this surrogate becomes tight as $\alpha \to 0$.
Our goal is to jointly optimize the beamforming vector and the AN vector to maximize the total secrecy, which leads 
to the problem formulation as
\begin{align}
\max_{\bm{W}, \bm{V}}\quad & R_{\mathrm{sum}} =\sum_{{k \in \mathcal{K}}} R_k^{\mathrm{sec}} \label{7}\\
\text{s.t.}\quad & \sum_{{k \in \mathcal{K}}} \| \bm{w}_k \|^2 + \sum_{{j \in \mathcal{J}}} \| \bm{v}_j \|^2 \leq 1, \label{eq8}
\end{align}
where \(\bm{W}\!=\![\bm{w}_1, \ldots, \bm{w}_K]\)\! is the beamforming matrix at the BS. The constraint in~(\ref{eq8}) ensures the total power of normalized beamforming and AN vectors remains within the unit budget. 

The problem is non-convex, and the beamforming and AN vectors are coupled to each other in the secrecy rate expression, further increasing the difficulty of direct optimization. Instead of using computationally intensive iterative solvers, we employ a channel-aware conditional diffusion model, which is a generative diffusion framework that directly learns the solution space from data. This approach supports fast sampling-based inference for new channel realizations, and can generate multiple high-quality candidate solutions, thereby enhancing real-time applicability in dynamic wireless environments. 
\begin{figure}[t] 
    \setlength{\abovecaptionskip}{-0.1cm}  
        \centering 
        \includegraphics[width=0.42\textwidth]{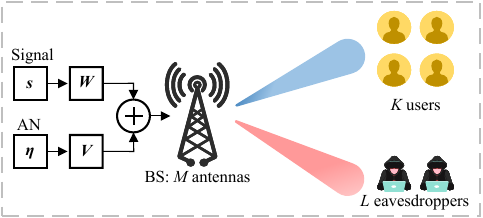} 
        \caption{Downlink MU-MISO wiretap system with joint beamforming and AN} 
        \label{Fig.main4} 
        \vspace{-0.4cm}
        \end{figure}

\section{Channel-Aware Conditional Diffusion Model}
In this section, we provide a detailed framework for the generation method based on the proposed diffusion model. This method learns the conditional 
distribution of the optimal strategy through model training. Once trained, the model can efficiently generate 
high-quality solutions.
\begin{figure*} 
    \centering 
    \includegraphics[width=0.78\textwidth]{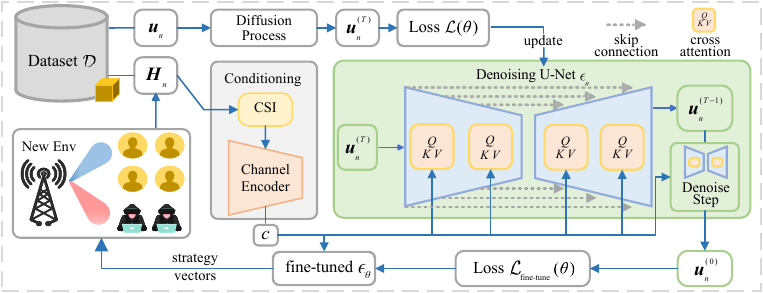}
    \caption{Channel-conditional diffusion model training and fine-tuning architecture for secure MU-MISO systems.} 
    \vspace{-0.3cm}
    \label{Fig_main} 
    \end{figure*}
\vspace{-0.5cm}
\subsection{Diffusion Model Interpretation}
We first define the composite channel \(\bm{h}\) as the concatenation of channel vectors from the BS to \(K\) users and \(L\) eavesdroppers, expressed as 
\begin{equation}
\setlength{\abovedisplayskip}{1pt}
\setlength{\belowdisplayskip}{2pt}
    \bm{h} = \left[ \bm{h}_1^{\top}, \ldots, \bm{h}_K^{\top}, \bm{h}_{\mathrm{E},1}^{\top}, \ldots, \bm{h}_{\mathrm{E},L}^{\top} \right]^{\top}
\in \mathbb{C}^{M(K+L)}.
\end{equation}
Similarly, we define the composite vector \(\bm{u}\) to encompass the transmit beamforming and AN design, expressed as
\begin{equation}
\setlength{\abovedisplayskip}{1pt}
\setlength{\belowdisplayskip}{2pt}
    \bm{u} = \left[ \bm{w}_1^{\top}, \ldots, \bm{w}_K^{\top}, \bm{v}_1^{\top}, \ldots, \bm{v}_J^{\top} \right]^{\top}
\in \mathbb{C}^{M(K+J)}.
\end{equation}
Now, consider the system model scenario with $N$ distinct channel realizations. For the $n$-th realization, we represent the composite channel and its corresponding optimal beamforming and AN vectors as \(\bm{h}_n\) and \(\bm{u}_n\) respectively. The collection $\mathcal{D}\!\!=\!\!\left\{ \left(\bm{h}_n,\bm{u}_n \right) \right\}_{n=1}^{N}$ serves as training dataset capturing the conditional distribution between \(\bm{h}_n\) and \(\bm{u}_n\). Due to the nonlinearity and high dimensionality inherent to secrecy-rate maximization, the conditional distribution \(P(\bm{u}_n\!\mid\!\bm{h}_n)\) is generally complex and multimodal. Instead of linearizing or simplifying the problem, our approach is to learn this complicated distribution directly using a diffusion model.

The architectural realization is illustrated in Fig.~\ref{Fig_main}. We first construct the dataset $\mathcal{D}$ through numerical simulation, where channel matrices $\bm{h}_n$ are sampled from standard wireless distributions, and corresponding near-optimal beamforming and AN vectors $\bm{u}_n$ are computed by employing the scheme proposed in~\cite{9973364}. Then, in order to condition the generative process, the complex CSI \(\bm{h}_n\) is processed into a real-valued, fixed-dimensional embedding \(\bm{c}_n\) via lightweight MLP, which is expressed as
\begin{equation}
\setlength{\abovedisplayskip}{1pt}
\setlength{\belowdisplayskip}{2pt}
    \bm{c}_n = \mathrm{MLP}\left(
\begin{bmatrix}
\Re(\bm{h}_n)^{\top},\ 
\Im(\bm{h}_n)^{\top}
\end{bmatrix}
\right)
\in \mathbb{R}^{d_c},
\end{equation}
where $d_c$ is the set embedding dimension. This uniform channel embedding ensures compatibility with the network input dimension, thus allowing the model to accommodate different antennas and user configurations. During training, the diffusion model gradually perturbs each observed beamforming and AN vector within a fixed $T$ steps and learns to reverse this perturbation. After training, the model can start from random Gaussian noise and iteratively denoise it to obtain feasible beamforming and AN design based on the new CSI input. This approach effectively learns the underlying mapping by modeling the entire conditional distribution rather than a single point solution, providing a diffusion model-based interpretation for the beamforming and AN design problem.

Specifically, a denoising U-Net serves as the backbone of the model, which iteratively refines the beamforming and AN vectors over $T$ steps. The U-Net follows an encoder-decoder structure of multiple downsampling and upsampling blocks. Skip connections link corresponding levels of the encoder and decoder, which enables the network to capture both global structure and local details. In our task, global structures refer to the vector pattern of beamforming and AN, indicating directions in which energy is concentrated or suppressed, while local details correspond to small variations in the amplitude and phase of the CSI. Moreover, we integrate CSI by introducing a cross-attention mechanism at the bottleneck layer and at multiple key locations of each downsampling and upsampling block. The current U-Net layer's features serve as queries, and the channel embedding provides keys and values. This enables the network to dynamically focus on distinct channel characteristics during denoising. For example, when the network needs to decide how to allocate power between beamforming and AN vector, the attention mechanism can automatically emphasize the feature dimensions associated with the eavesdropping channel information, thus ensuring that the AN is efficiently injected into the appropriate subspace. This dynamic characteristic modulation based on CSI is the key of conditional generation.

Based on the preceding description, we reformulate the optimization problem as a conditional generation task guided by CSI. In contrast to conventional generative models applied to wireless communication tasks, our diffusion-based approach is specifically designed for the transmission strategy optimization of PLS, offering a data-driven alternative to computationally intensive non-convex optimization methods.
\vspace{-0.3cm}
\subsection{Conditional Diffusion Process}
\vspace{-0.1cm}
To implement this diffusion model, we transform the input $\bm{u}_n$ into a two-dimensional real-valued format, explicitly separating real and imaginary components. 
Specifically, $\bm{u}_n$ is expressed as
\begin{equation}
    \setlength{\abovedisplayskip}{1pt}
    \setlength{\belowdisplayskip}{2pt}
    \bm{z}_n = \left[
    \Re(\bm{u}_n)^{\top},\ 
    \Im(\bm{u}_n)^{\top}
    \right]^{\top}
    \in \mathbb{R}^{2M(K+J)}.
    \end{equation}
This representation enables the network to share parameters across antennas, effectively capturing correlated behaviors inherent in spatial beamforming and AN design. 

Then, we adopt a channel-conditional denoising diffusion probabilistic model (DDPM) as the backbone~\cite{NEURIPS2020_4c5bcfec}. 
The forward diffusion process perturbs the data distribution 
by adding noise, and the reverse denoising process, parameterized by a neural network, iteratively removes noise to reconstruct samples from the target data distribution. 
We specify a variance schedule $\{\beta_1, \beta_2, \dots, \beta_T\}$ 
where $0<\beta_t<1$. At each step $t$, we construct a Markov chain by adding Gaussian noise with variance $\beta_t$ to the 
signal. Specifically, the forward transition is defined as 
\begin{equation}
\setlength{\abovedisplayskip}{1pt}
\setlength{\belowdisplayskip}{2pt}
q\left(\bm{z}_n^{(t)} \mid \bm{z}_n^{(t-1)}\right) = \mathcal{N} \left( \bm{z}_n^{(t)};\sqrt{1 - \beta_t} \, \bm{z}_n^{(t-1)}, \, \beta_t \bm{I} \right),
\label{e11}
\end{equation}
where \(\bm{I}\) is the identity matrix. By iterating the formula~(\ref{e11}), we can derive a closed-form expression representing \(\bm{z}_n^{(t)}\) in terms of 
the original sample \(\bm{z}_n\) and the accumulated noise. Define $\alpha_t = 1-\beta_t$ and $\bar{\alpha}_t = \prod_{s=1}^t \alpha_s$, the expression can be written as
\begin{equation}
    \setlength{\abovedisplayskip}{1pt}
    \setlength{\belowdisplayskip}{1pt}
\bm{z}_n^{(t)} = \sqrt{\bar{\alpha}_t} \, \bm{z}_n + \sqrt{1 - \bar{\alpha}_t} \, \boldsymbol{\epsilon}, 
\label{e12}
\end{equation}
where $\boldsymbol{\epsilon}$ is a standard Gaussian noise independent for each $t$. After $T$ steps, \(\bm{z}_n^{(T)}\) obeys an isotropic Gaussian distribution. 
Channel condition \(\bm{c}_n\) has no effect on the forward process~\cite{10557650}, and is only introduced in the reverse process.

The reverse denoising process begins by sampling \(\bm{z}_n^{(T)} \sim \mathcal{N} \left(0,\bm{I}\right)\) and iteratively applies the conditional 
transition
\begin{equation}
\small
    \setlength{\abovedisplayskip}{2pt}
    \setlength{\belowdisplayskip}{2pt}
p_{\bm{\Theta}}\left(\bm{z}_n^{\!(t\!-\!1)} \!\mid \!\bm{z}_n^{(t)}, \!\bm{c}_n\right) 
\!=\! \mathcal{N}\!\left(\bm{z}_n^{(t-1)}; \boldsymbol{\mu}_{\bm{\Theta}}(\!\bm{z}_n^{(t)}, t, \bm{c}_n), \!\boldsymbol{\Sigma}_{\bm{\Theta}}(\bm{z}_n^{\!(t)}, t) \right),
\end{equation}
progressively reconstructing the target strategy vector. Here, $\boldsymbol{\mu}_\Theta(\bm{z}_n^{(t)}, t, \bm{c}_n)$ is a learned denoising function 
output by neural network denoiser. In implementation, we train a neural network 
$\boldsymbol{\epsilon}_{\bm{\Theta}}(\bm{z}_n^{(t)}, t, \bm{c}_n)$ to predict the Gaussian noise $\boldsymbol{\epsilon}$ in~(\ref{e12}). This noise prediction network takes the current noise vector $\bm{z}_n^{(t)}$, time step $t$ and channel $\bm{c}_n$ as inputs, and outputs the noise vector with the same dimension as $\bm{z}_n^{(t)}$. With the learned noise estimator $\boldsymbol{\epsilon}_{\bm{\Theta}}$, the posterior mean is
\begin{equation}
\small
\setlength{\abovedisplayskip}{2pt}
\setlength{\belowdisplayskip}{2pt}
\boldsymbol{\mu}_{\bm{\Theta}}(\bm{z}_n^{(t)}, t, \bm{c}_n)
= \!\frac{1}{\sqrt{\alpha_t}}\! \left( \bm{z}_n^{(t)} 
- \frac{1 - \alpha_t}{\sqrt{1\! - \!\bar{\alpha}_t}} 
\, \boldsymbol{\epsilon}_{\bm{\Theta}}(\bm{z}_n^{(t)}, t, \bm{c}_n)\right),
\label{e14}
\end{equation}
and we employ this parameterized form to implement the steps of the reverse diffusion iteration. The procedure supports batched sampling, 
so multiple solutions can be generated in parallel without additional runtime. Incorporating DDIM further cuts the number of denoising steps, markedly improving sampling efficiency~\cite{song2022denoisingdiffusionimplicitmodels}.
After multiple iterations, the model progressively denoises \(\bm{z}_n^{(T)} \sim \mathcal{N} \left(0,\bm{I}\right)\) into a sample
\(\bm{z}_n^{(0)}\). By construction, the $\bm{z}_n^{(0)}$ should resemble a plausible optimized strategy for the channel $\bm{c}_n$. Finally, re-assembling the real-imaginary channels yields the complex strategy vector
\begin{equation}
\bm{u}_n^{(0)} = \mathrm{unvec}\left(\Re\{\bm{z}_n^{(0)}\} + \mathrm{j}\Im\{\bm{z}_n^{(0)}\}\right) \in \mathbb{C}^{M(K+J)}.
\end{equation}

The neural network parameters ${\bm{\Theta}}$ are optimized so that the reverse diffusion model yields samples $\bm{u}_n^{(0)}$ that follow the conditional data distribution \(P(\bm{u}_n\!\mid\!\bm{h}_n)\).
\vspace{-0.3cm}
\subsection{Model Training and Fine-Tuning}
Prior to model training, we standardize both the channel and strategy vectors in dataset \(\mathcal{D}\) through feature-wise standardization, ensuring uniform scaling across all input dimensions. This helps the model concentrate on learning potential secure transmission principles and promotes more stable convergence of training. Then, the model parameters are learned by optimizing a denoising objective that captures the essential characteristics of the diffusion process. Specifically, we minimize the following loss function
\begin{equation}
    \setlength{\abovedisplayskip}{3pt}
\setlength{\belowdisplayskip}{3pt}
\mathcal{L}({\bm{\Theta}}) 
= \mathbb{E}_{t, \bm{z}_n, \boldsymbol{\epsilon}} \!
\left[\! \left\| \boldsymbol{\epsilon} \!
- \!\boldsymbol{\epsilon}_{\bm{\Theta}}\!\!\left( \!
\sqrt{\bar{\alpha}_t} \, \bm{z}_n \!+ \!\sqrt{1 \!- \!\bar{\alpha}_t} \, \boldsymbol{\epsilon}, \, t, \bm{c}_n
\right) \!\right\|^2 \!\right].
\label{e13}
\end{equation} 
The training process computes the mean squared error (MSE) between the predicted and actual noise, enabling the model to learn the data distribution through iterative denoising.
\begin{figure}[t]
    \centering
    \setlength{\abovecaptionskip}{0.cm}
    \subfigure[Convergence of model training.]{
    \label{fig_4a}
    \includegraphics[scale=0.45]{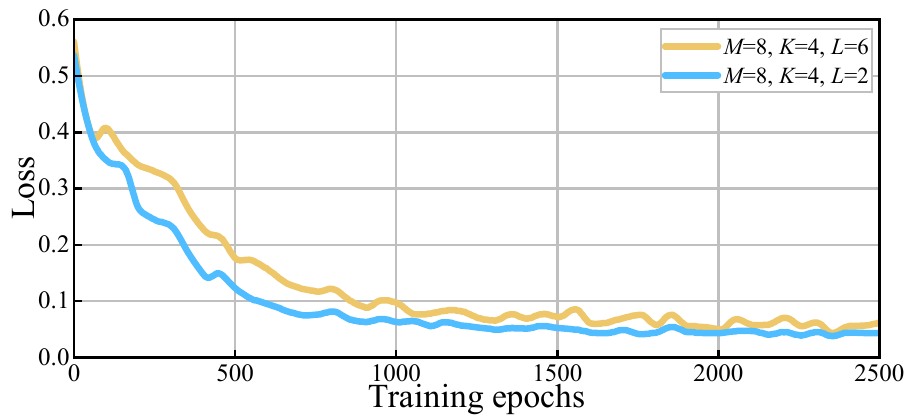}}\\[-1ex]
    \subfigure[Convergence of model fine-tuning.]{
    \label{fig_4b}
    \includegraphics[scale=0.45]{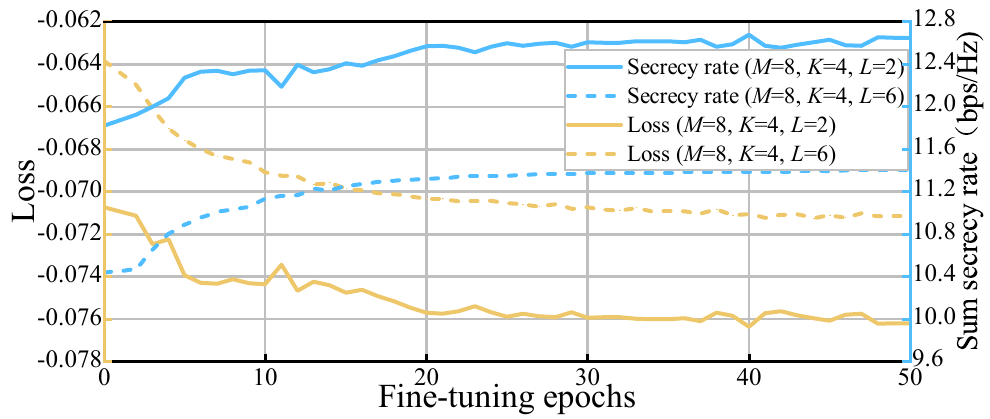}}
    \caption{Convergence of model training and fine-tuning.}
    \vspace{-0.4cm}
    \end{figure}

    \begin{figure*}[htbp]
        \centering 
        \subfigure[Performance with respect to transmit power.]{
        \label{fig_3a}
        \includegraphics[width=0.33\textwidth]{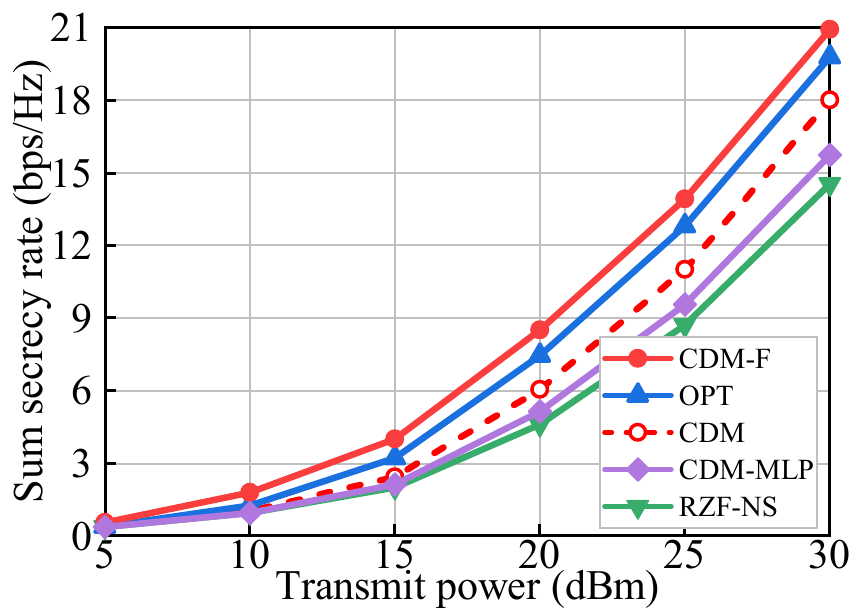}}
        \subfigure[Performance with respect to number of Eves.]{
        \label{fig_3b}
        \includegraphics[width=0.322\textwidth]{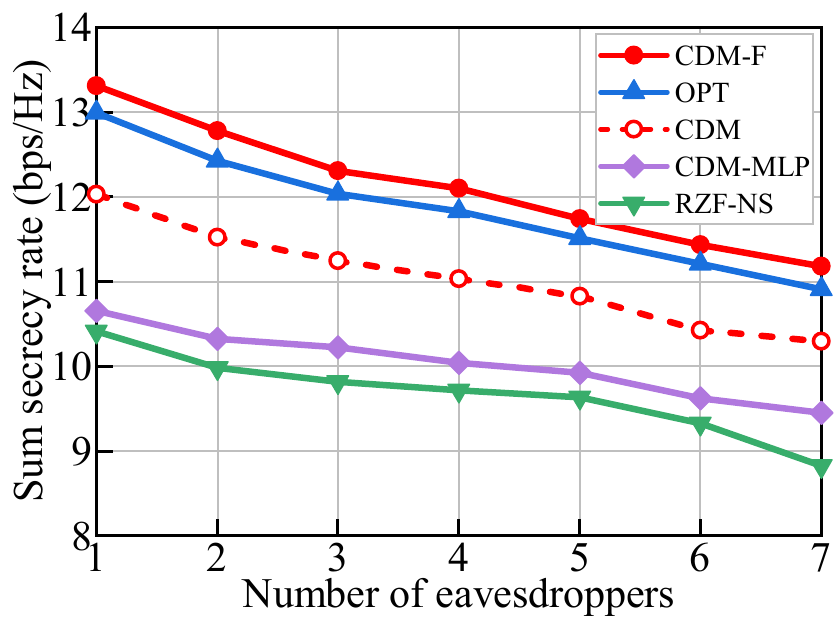}}
        \subfigure[Performance with respect to number of users.]{
        \label{fig_3c}
        \includegraphics[width=0.33\textwidth]{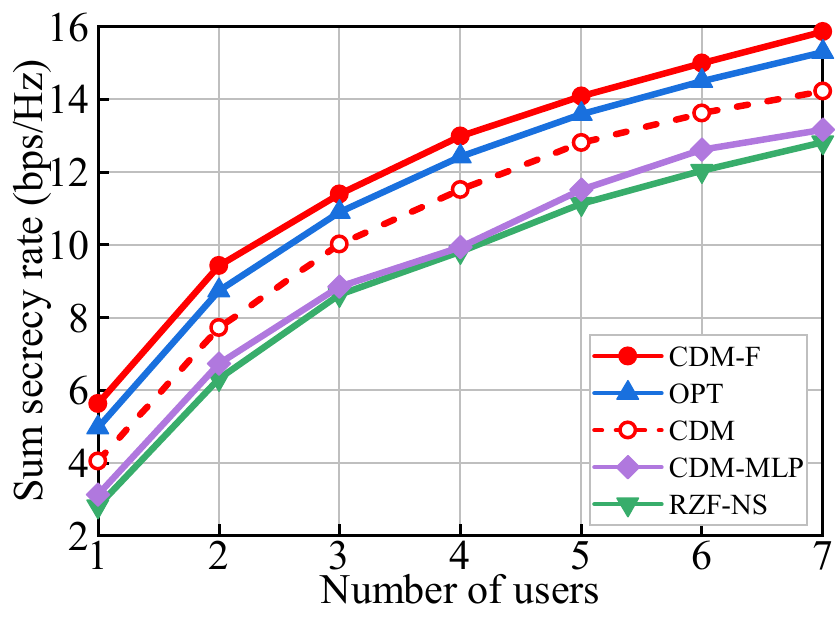}}
        \subfigure[Performance with respect to noise power.]{
        \label{fig_3d}
        \includegraphics[width=0.33\textwidth]{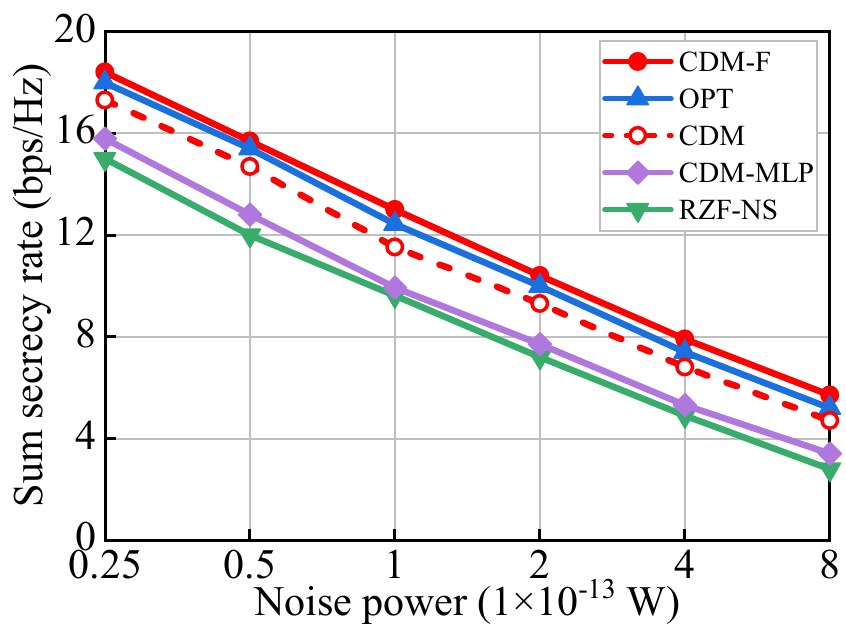}}
        \vspace{-0.3cm}
        \caption{Performance under different algorithms.}
        \vspace{-0.3cm}
        \label{Fig.main}
        \end{figure*}
While the trained model has effectively learned a general approximation of the near-optimal solution distribution, minor 
deviations from the true optimal distribution persist. To further enhance performance, we conduct targeted fine-tuning guided by 
the original optimization objective of maximizing the sum secrecy rate. Specifically, for a given set of channel conditions $\bm{h}_n$, the pre-trained 
diffusion model generates candidate solutions $\bm{u}_n^{(0)}$ and calculates the corresponding sum secrecy rate via~(\ref{eq6}). 
Applying the penalty term to ensure the power constraint, the fine-tuning loss function is defined as
    \begin{equation}
        \small
        \setlength{\abovedisplayskip}{2pt}
        \setlength{\belowdisplayskip}{2pt}
    \mathcal{L}_{\text{fine-tune}}({\bm{\Theta}}) 
    = - \mathbb{E} \left[ {R}_{\text{sum}}(\bm{u}_n^{(0)}, \bm{h}_n) \right] 
    + \lambda \left( \left\| \bm{u}_n^{(0)} \right\|_2 - \delta \right)^2,
    \label{e17}
    \end{equation}
where $\delta$ is the target norm introduced to enforce the power constraint in~(\ref{eq8}), and $\lambda$ serves as a hyperparameter which balances the numerical scale between the sum secrecy rate objective and the power constraint penalty. Since the model has already explored the neighborhood close to the optimal solution during the initial training process, this fine-tuning process effectively narrows the distribution and concentrates it around higher-secrecy-rate regions of the solution space, thereby further improving the performance.

Furthermore, the computational complexity of the proposed generative framework is analyzed as follows. Let $\mathcal{I}$ denote the number of encoder and decoder blocks in the U-Net, and at each block $i \in \{1, \dots, \mathcal{I}\}$, let $S_i$ and $C_i$ denote the spatial sequence length and channel dimension at block $i$, respectively. The complexity per timestep decomposes into convolutional and attention terms: the former contains $N_{\mathrm{conv},i}$ residual blocks, and the complexity of each block is $\mathcal{O}(S_i C_i^2 K_{\mathrm{ker},i}^2)$ with kernel size $K_{\mathrm{ker},i}$; the latter involves $N_{\mathrm{attn},i}$ cross-attention modules, and each with a complexity of $\mathcal{O}(S_i^2 C_i + S_i C_i^2)$. Consequently, with $E$ training epochs and a dataset of size $N$, the total training complexity is $\mathcal{O}\big(E N \sum_{i=1}^{\mathcal{I}} [N_{\mathrm{conv},i} S_i C_i^2 K_{\mathrm{ker},i}^2 + N_{\mathrm{attn},i}(S_i^2 C_i + S_i C_i^2)]\big)$. During the inference stage, denoting the number of steps in DDIM sampling as $T_\mathrm{ddim}$, the complexity of inference is $\mathcal{O}\big(T_\mathrm{ddim} \sum_{i=1}^{\mathcal{I}} [N_{\mathrm{conv},i} S_i C_i^2 K_{\mathrm{ker},i}^2 + N_{\mathrm{attn},i}(S_i^2 C_i + S_i C_i^2)]\big)$.

\section{Simulation Results}

To evaluate the proposed scheme, we simulate an isolated single-cell downlink served by an 8-antenna BS.  Users are randomly distributed 50 m - 80 m from the BS, with each eavesdropper placed within 5 m of a random user. Path loss is modeled by $\mathcal{L} = \mathcal{L}_0 - 10 \, \xi \log_{10} \left( \frac{d_1}{d_0} \right)$ dB, with reference loss $\mathcal{L}_0 = -30$, path-loss exponent $\xi = 2.2$ and link distance $d_1$. Small-scale fading is assumed to follow independent Rayleigh statistics. Unless stated otherwise, the BS transmit power is 20 dBm, and the noise power is -100 dBm. We set $\alpha = 0.01$, $\delta = 1$ and $\lambda = 0.001$. The network is trained for \( T = 1000 \) steps with \( \beta_t \in \left[10^{-4},\, 0.02\right] \). During the inference stage, denoising is performed using DDIM with $T_\mathrm{ddim}=50$. The channel embedding dimension is $d_c=256$ , and the U-Net consists of $\mathcal{I}=3$ encoder and decoder blocks, with the spatial sequence length $S_i$ progressing through 64$\rightarrow$16$\rightarrow$4 and the channel dimension $C_i$ transitioning as 64$\rightarrow$128$\rightarrow$256 across the respective blocks. For all blocks, $N_{\mathrm{conv},i}=2$, $N_{\mathrm{attn},i}=2$ and $K_{\mathrm{ker},i}=3$. Model parameters are optimized with Adam at a learning rate of \( 10^{-4} \).

We first show the convergence of the proposed conditional generation model during training.  As shown in Fig.~\ref{fig_4a}, 
the loss of the model in the first stage decreases gradually with the number of iterations and converges overall after 
about 1000 epochs, indicating that the model has learned to accurately predict the noise involved in the total time 
steps $T=1000$.  After that, although the loss curve stabilized, it continued to decrease slightly, indicating that the 
model continued to learn the high-frequency details and residual errors of the data distribution in the subsequent stage. 
The results demonstrate the effectiveness of 
the proposed generative framework in noise prediction and distribution modeling. Moreover, Fig.~\ref{fig_4b} shows the 
convergence process in the fine-tuning stage. Compared with the first-stage training, the fine-tuning leveled off after 
about 40 epochs, indicating that targeted fine-tuning can quickly discover a better strategy in very few additional 
training rounds.  At the same time, the fine-tuning loss curve directly reflects the potential of the model to further 
improve the confidentiality rate target, and verifies the effective gain in security performance of the method.

For performance evaluation, we compare the following schemes: 1) CDM: the proposed generative approach, which directly generates joint beamforming and AN vectors conditioned on CSI. 2) CDM-F: an enhanced variant of CDM obtained by further fine-tuning with explicit secrecy-rate optimization. 3) OPT: joint generalized power iterative precoding, a traditional iterative optimization-based beamforming and AN design~\cite{9973364}. 4) RZF-NS: regularized zero-forcing beamforming combined with null-space AN. 5) CDM-MLP: a modified variant of CDM obtained by replacing the U-Net backbone with a multilayer perceptron.

Fig.~\ref{fig_3a} demonstrates the secrecy rate variation versus transmit power for a system configuration with $K=2$, and $L=4$. Both CDM and CDM-F outperform RZF-NS, with CDM approaching OPT performance at medium-high power, indicating successful learning of near-optimal strategies. CDM-F achieves the highest rates across all power levels, which verifies the effectiveness of secrecy-rate-guided fine-tuning in exploring better strategies. The proposed CDM and CDM-F outperform compared to the CDM-MLP. This performance gap highlights the critical role of the U-Net architecture. The convolutional structure and multi-scale feature extraction of U-Net are more effective in handling the inherent spatial correlations in antennas and beamforming vectors. In contrast, the replaced MLP usually has difficulty simultaneously considering both local high frequencies and global low frequencies while maintaining a similar complexity. Fig.~\ref{fig_3b} shows the change of the system secrecy rate with the increase of the number of eavesdroppers under the conditions $K=4$, and $P=20$ dBm. The secrecy rate of CDM-F has been improved to a certain extent throughout the entire process compared to OPT, which confirms that fine-tuning enhances the model's implicit interference cancellation strategy. Meanwhile, CDM consistently surpasses RZF-NS, indicating that the learned AN injection is more effective in adapting to multiple wiretap channels than the fixed null-space design. This shows the generative model has learned how to effectively counter the multi-channel interference of multiple eavesdroppers in training. Fig.~\ref{fig_3c} explores the scalability of multi-user by increasing the number of users under the conditions of $L=2$, and $P=20$ dBm. All curves increase almost linearly because additional users bring extra spatial degrees of freedom that can be exploited by beamforming. CDM-F maintains optimal performance and CDM shows stable behavior across varying user configurations, indicating its ability to capture essential strategy characteristics. Fig.~\ref{fig_3d} evaluates the secrecy rate performance across different noise power levels, with the conditions of $K=4$, $L=2$ and $P=20$ dBm. The proposed CDM-F and CDM consistently maintain performance superior to RZF-NS and CDM-MLP baselines. By capturing the inherent non-linear mapping between channel conditions and the optimal beamforming, this model demonstrates zero-shot generalization ability in this regard, enabling it to generalize directly to varying noise environments without requiring new parameter tuning.

\section{Conclusion}
In this paper, we have proposed a generative diffusion model-based framework for joint beamforming and AN design in MU-MISO wiretap channels. 
Our proposed method leverages the powerful distribution-learning capability of diffusion models to reformulate the complex, non-convex optimization problem into a conditional generative task guided by CSI. 
Through training and targeted fine-tuning, the diffusion model effectively learns to generate high-quality transmission strategies directly from noise under various channel conditions. Simulation results have shown that our proposed method achieves higher secrecy rates compared to traditional benchmarks, 
highlighting its potential for real-time PLS applications in dynamic wireless communication scenarios.

\bibliographystyle{IEEEtran}
\bibliography{reference}

\end{document}